\documentclass{an}
\usepackage{graphicx}
\usepackage{psfig}
\usepackage{times}
\usepackage{fancyhdr}
\begin{document}

\title{Widely separated binary systems of very low mass stars}

\author{N.~Phan-Bao
       \inst{1}
       \and
       E.L.~Mart\'{\i}n
       \inst{2,3}
       \and
       C.~Reyl\'e
       \inst{4}
       \and
       T.~Forveille
       \inst{5,6}
       \and
       J.~Lim
       \inst{1}}
\institute{Institute of Astronomy and Astrophysics, Academia Sinica.
           P.O. Box 23-141, Taipei 106, Taiwan, R.O.C.
          \and
	  Instituto de Astrof\'{\i}sica de Canarias, C/ V\'{\i}a L\'actea  
          s/n, E-38200 La Laguna (Tenerife), Spain.
          \and
          University of Central Florida, Dept. of Physics, PO Box 162385, 
          Orlando, FL 32816-2385, USA.
          \and
	  CNRS UMR6091, Observatoire de Besan\c{c}on, BP1615, 
	  25010 Besan\c{c}on Cedex, France.
	  \and
	  Canada-France-Hawaii Telescope Corporation, 65-1238 Mamalahoa 
          Highway, Kamuela, HI 96743 USA.	
	  \and
	  Laboratoire d'Astrophysique de Grenoble, Universit\'e J. 
          Fourier, B.P. 53, F-38041 Grenoble, France.}
\date{Received; accepted; published online}

\abstract{In this paper we review some recent detections of wide binary brown dwarf systems 
and discuss them in the context of the multiplicity properties of very low-mass stars
and brown dwarfs.
}

\correspondence{pbngoc@asiaa.sinica.edu.tw}

\maketitle

\section{Introduction}
Binary systems have been studied for decades to measure accurate stellar
masses, and to test evolutionary models and star formation theories.
Considerable attention has recently been paid to very low-mass (VLM) 
binaries in the solar neighborhood 
(Mart\'\i n et al. \cite{martin99a}; Close et al. \cite{close02}, 
\cite{close03}; Bouy et al. \cite{bouy}; Burgasser et al. \cite{burgasser03}; 
Forveille et al. \cite{forveille05}),
as well as in nearby young open clusters and associations
(Mart\'\i n et al. \cite{martin03}; Chauvin et al. \cite{chauvin}).

The properties of VLM binaries are an important constraint for models 
of star-formation and evolution. It has been debated in the literature 
whether the properties of VLM binaries and stellar binaries differ, 
implying different formation mechanisms (Kroupa et al. \cite{kroupa03}),
or whether the binary properties instead show continuous trends with 
decreasing primary mass, implying that VLM binaries form through the same
processes as stellar binaries (Luhman \cite{luhman04b}). Clearly there is 
a need for a larger sample of observed VLM binaries, particularly at wide
separations where few of them are known. 
One leading model of brown dwarf (BD) formation
is that they form and are ejected in unstable multiple systems within
small clusters. Since the ejection models (Reipurth 
\& Clarke \cite{reipurth}; Bate et al. \cite{bate02}) suggest that the
binary BD systems that do exist must be close (separations~$\leq$~10~AU).
The detection of wide VLM binary systems has thus become an important 
test of the ejection models. The first wide binary BDs have been found in 
young ($<$10~Myr) associations or clusters (Luhman \cite{luhman04a}; Chauvin et al. \cite{chauvin}).
Very recently Phan-Bao et al. (\cite{phan-bao05}) 
found a 33~AU ultracool binary system, Bill\`eres et al. (\cite{billeres}) have also
reported a discovery of a 200~AU M8.0+L0 binary, both of them in the field and their detection
based on the DENIS survey.

\section{LP~714-37AB: the closest wide binary system of very low mass stars}

In Phan-Bao et al. ({\cite{phan-bao05}) we reported the discovery of a new wide
binary system of very low mass stars with projected separation of 33~AU.
The calibration of the PC3 index 
to spectral type (Mart\'{\i}n et al. \cite{martin99b}) gives spectral types of M5.5 for component A and M7.5 
for component B, with an uncertainty of $\pm$0.5 subclass. 
Figure~\ref{fig_pm} shows that there is no background star at the position
of the system in either the SERC-I image ($I$ band) or the DENIS-I image,
and therefore demonstrates that the system LP~714-37 is a physical binary.
The properties of this system are given in Table~\ref{table_system}.

\begin{figure}[t]
\hspace{1.2cm}
\psfig{width=6.0cm,file=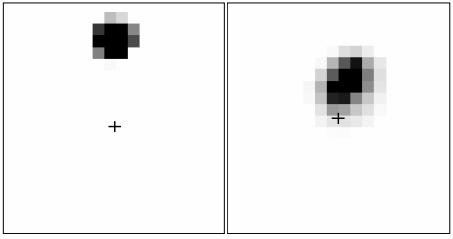,angle=0}
\caption{Archival images of LP 714-37: SERC-I (left, epoch: 1986.940) and
DENIS-I (right, epoch: 2000.896). The cross indicates the position of 
component B at the 2003.915 epoch of the ESO-New Technology Telescope image. 
Clearly, with the proper motion of the system ($\mu_{\rm \alpha}$ = $-$117~mas/yr and 
$\mu_{\rm \delta}$ = $-$382~mas/yr, Phan-Bao et al. \cite{phan-bao03}),
the SERC-I image would easily separate the two objects if they were not physically 
associated. The size of each image is 20$\times$20\arcsec, and North is 
up and East to the left.}
\label{fig_pm}
\end{figure}
%

Using the PC3 index to magnitudes relation given in Crifo et al. (\cite{crifo})
and comparison with the DENIS apparent magnitudes give the distance only at 18~pc for
our system, this makes LP~714-37AB the closest wide utracool binary 
system (separation $>$ 30~AU) in the field up to now.
Table~\ref{table_review} lists all known wide ultracool field binary dwarfs.

\begin{table}
   \caption{Summary of the LP~714-37AB binary system}
    \label{table_system}
  $$
   \begin{tabular}{llllll}
   \hline 
   \hline
   \noalign{\smallskip}
Stars               & PC3  & SpT &  Mass        &  Sep.  &  dist. \\
                    &      &     & (M$_{\odot}$)& (AU)   &  (pc)\\           
  (1)               &(2)   &(3)  & (4)          & (5)    &  (6)\\
    \noalign{\smallskip}
    \hline 
LP 714-37A &  1.32 & M5.5  &  0.11 &   33.1$\pm$4.0 &  18.1$\pm$2.2  \\
LP 714-37B &  1.72 & M7.5  &  0.09 &                &                \\
    \noalign{\smallskip}
    \hline 
   \end{tabular}
  $$
  \begin{list}{}{}
  \item[] 
{\it Column 1}: NLTT name.
{\it Columns 2} \& {\it 3}: The PC3 index and spectral types derived from the (PC3, spectral type)
relation of Mart\'{\i}n et al. (\cite{martin99b}).
{\it Column 4}: Mass determinations for 1-5~Gyr from the models of Baraffe et al. (\cite{baraffe98}).
{\it Column 5}: Projected separation.
{\it Column 6}: Spectrophotometric distance.
  \end{list}
\end{table}
%

\section{Discussion}

Recent surveys demonstrate that VLM binaries with large separations 
($>$~30~AU) are rare in the field, but can be found in young associations and 
clusters: 2MASS~J1101-7732 
(240~AU separation, in Chamaeleon~I, Luhman \cite{luhman04a}); 
2MASS~J1207-3932 (55~AU, in TW Hydrae, Chauvin et al. \cite{chauvin}).
Phan-Bao et al. (\cite{phan-bao05}) reported the discovery of 
a 33~AU VLM binary, Mart\'{\i}n et al. (\cite{martin00}) found that
CFHT-Pl-18 is a 35~AU VLM binary; and very recently, Bill\`eres et al. ({\cite{billeres})
discovered an M8.5+L0 pair (DENIS-P~J055146.0-443412.2) with a physical separation over 200~AU,
all of them in the field. 

Since ejection models suggest that the binary DB systems that do exist
must be close. 
And therefore the existence of the extremely wide binaries (e.g., 200~AU separation) becomes 
a very important test of the ejection models (Bate \& Bonnell \cite{bate05}).
One should note however that the numerical models to date suffer from small number statistics. 
One should also note that the relevant quantity is the total mass of the
system, and that the apparent binaries could possibly be triple or 
higher order multiple systems, with a correspondingly higher total mass.
This would make them analogs of the GJ1245ABC triple system, which
consists of two M5.5 and one $\sim$M8 dwarfs with separations of 32 and 5 AUs.
Triple systems could thus potentially explain the apparent excess of
wide VLM binaries, and adaptive optics imaging of LP~714-37, DENIS-P~J055146.0-443412.2
would be of obvious interest to clarify its true multiplicity.
\begin{table}
   \caption{Widely separated binary systems (separations $>$30~AU) in the field}
    \label{table_review}
  $$
   \begin{tabular}{lllll}
   \hline 
   \hline
   \noalign{\smallskip}
Stars               & Sep.     & SpT$_{\rm A}$/SpT$_{\rm B}$ &  Dist.  & Ref.    \\
                    & (AU)     &                       & (pc)    &    \\           
  (1)               &(2)       &(3)                    & (4)     &  (5)   \\
    \noalign{\smallskip}
    \hline 
LP 714-37            &~\,33 & M5.5/M7.5  &    ~\,18        &  ~\,1       \\
CFHT-Pl-18           &~\,35 & M8.0/M8.0  &    105       &  2,3     \\
DENIS-P J0551$-$4434 &  200 & M8.5/L0.0  &    100       &  ~\,4       \\
    \noalign{\smallskip}
    \hline 
   \end{tabular}
  $$
  \begin{list}{}{}
  \item[] 
{\it Column 1}: Star name.
{\it Columns 2} \& {\it 3}: Projected separation and spectral types of components.
{\it Column 4}: Distance estimate.
{\it Column 5}: References: (1) Phan-Bao et al. (\cite{phan-bao05}); (2) Mart\'{\i}n et al. (\cite{martin00});
(3) Bouy et al. (\cite{bouy}); (4) Bill\`eres et al. (\cite{billeres}).
  \end{list}
\end{table}

\acknowledgements
P-B.N. is grateful to the DENIS consortium for access to 
the DENIS data used by his very low mass stars search. Partial funding was provided by NSF grant AST 02-05862.
P-B.N. would like to thank the LOC of the La Palma workshop on Ultra-Low Mass Star Formation
for organizing a very enjoyable conference.

\end{document}